\def\year{2020}\relax
\documentclass[letterpaper]{article} 
\usepackage{aaai20}  
\usepackage{times}  
\usepackage{helvet} 
\usepackage{courier}  
\usepackage[hyphens]{url}  
\usepackage{graphicx} 
\usepackage{graphicx}  
\usepackage{latexsym}
\usepackage{booktabs} 
\usepackage{footnote}
\makesavenoteenv{tabular}
\makesavenoteenv{table}

\usepackage{amsmath}
\usepackage{amssymb}
\usepackage{mathrsfs}
\usepackage{makecell}
\usepackage[]{algorithm2e}
\usepackage{graphicx}
\usepackage{verbatim}
\usepackage{bm}
\usepackage{arydshln}
\usepackage{ dsfont }

\title{Corpus-Level End-to-End Exploration  for Interactive Systems }

\author{Zhiwen Tang\ and Grace Hui Yang\\
InfoSense, Department of Computer Science\\
Georgetown University\\
zt79@georgetown.edu  \    huiyang@cs.georgetown.edu
}

\begin{document}

\maketitle

\begin{abstract}
A core interest in building Artificial Intelligence (AI) agents is to let them interact with and assist humans. 
One example is Dynamic Search (DS), which models the process that a human  works with a search engine agent to accomplish a complex and goal-oriented task.  
Early DS agents using Reinforcement Learning (RL) have only achieved limited success for (1) their lack of direct control over which documents to return and (2) the difficulty to recover from wrong search trajectories. In this paper, we present a novel corpus-level end-to-end exploration (CE3) method to address these issues. In our method, 
an entire text corpus is compressed into a global low-dimensional  representation, which enables the agent to gain access to the full state and action spaces, including the under-explored areas. We also propose a new form of retrieval function, whose linear approximation allows end-to-end manipulation of documents. Experiments on the Text REtrieval Conference (TREC) Dynamic Domain (DD) Track show that CE3 outperforms the state-of-the-art DS systems. 
\end{abstract}

\section{Introduction}

Retrieval-based interactive systems, including multi-turn Question Answering (QA) \cite{DBLP:journals/tacl/ReddyCM19}, dialogue systems \cite{DBLP:conf/acl/SankarSPCB19}, and dynamic search systems  \cite{yang2016dynamic}, study the interaction between a human user  and an intelligent agent when they work together to accomplish a goal-oriented task. 
Reinforcement Learning (RL) becomes a natural solution to these interactive systems \cite{yang2016dynamic,li2016multiple,Hu:2018:RLR:3219819.3219846} for its emphasis on adaptation and exploration. Prior work on this topic has investigated bandits-based \cite{li2016multiple}, value-based \cite{Luo:2014:WSD:2600428.2609629}, and policy-based \cite{Hu:2018:RLR:3219819.3219846} RL methods. In these approaches, oftentimes, a repository of documents (or knowledge) and inputs from a human user, are treated as the learning environment for the AI agent; and the agent's actions usually take two steps -- first, reformulating queries (or questions) based on user responses; second, retrieving relevant information to fulfill those queries via some off-the-shelf retrieval tools. Such a pipeline is a convenient use of existing Information Retrieval (IR) techniques; however, it comes with a few drawbacks.  

First, most existing retrieval functions are optimized over precision at top ranks, which is demanded by the limited cognitive load a human user could afford when examining the results. This bias is engraved in all ready-to-use retrieval tools. The consequence is that results that are good but not as optimal would have little chance to show up. It might be ideal when there is only one run of retrieval, such as in single-turn QA or ad-hoc document retrieval. In multi-turn interactions, however, early triage of lower-ranked documents would lead to long-term loss that can not be easily recovered. Classic works on exploratory search \cite{white2009exploratory} and information seeking \cite{Marchionini:2006:ESF:1121949.1121979} named this phenomenon  ``berry picking" --  which depicts a tortuous search trajectory where only very limited useful information can be obtained at each single step because the search space is so restricted by the top results. This makes the RL agent's learning very challenging because the agent is not able to ``explicitly consider the {\it whole} problem of a goal-oriented"  process \cite{Sutton1998}. 

Second, widely-used retrieval functions, including TF-IDF  \cite{DBLP:journals/ipm/SaltonB88} and BM25 \cite{bm25}, are  non-differentiable. Common functions to reformulate queries \cite{huang2009analyzing} are non-differentiable, too. These non-differentiable functions prevent a gradient-based RL method from updating its gradient through; thus a user's feedback would not have real control over which documents to return. Consequently, the retrieval results could look random.  Nonetheless, these functions remain quite popular due to their readiness. 

In this paper,  using dynamic search as an illustrating example, we propose a different solution to retrieval-based interactive agents. In our corpus-level end-to-end exploration algorithm (CE3), at each time step,  we compress a text corpus into a single global representation and  used  to support the agent's full  exploration in the state and action spaces. In addition, 
 we propose  a novel   differentiable retrieval function that allows an RL agent to directly manipulate documents. 
Experiments on the Text REtrieval Conference (TREC) Dynamic Domain 2017 Tracks demonstrate that our method significantly outperforms previous DS 
approaches. It also shows that our method is able to quickly adjust the search trajectories and recover from losses in early interactions. Given the fundamental issues it addresses,  we believe CE3's success can be extended to other interactive AI systems if they access information using retrieval functions.

\section{Related Work} \label{related_work}

The work closest to ours is perhaps KB-InfoBot \cite{P17-1045}. It is a dialogue system to find movies  from a large movie knowledge base (KB).  Similar to us, KB-InfoBot used a global representation for {\it all} movies in its database to represent the states. To do so, it estimated a global distribution over the entire set of movie entities, conditioned on user utterances.
The distribution was fed into a deep neural network to learn the agent's action. Also similar to us, KB-InfoBot used a differentiable lookup function to support end-to-end manipulation of data entities. 
The two works differ in that their dialogue agent ran on a structured database and completed a task by iteratively filling the missing slots, while ours is for unstructured free text and accomplishes a task by iteratively retrieving documents that is relevant to the search task.

Knowing a global model that oversees the entire text collection  has shown to be  beneficial to retrieval in conventional IR research. For instance, \citeauthor{liu2004cluster} proposed to develop corpus-level clusters by K-means and then used them to smooth out multinomial language models. \citeauthor{wei2006lda} also used Latent Dirichlet Allocation (LDA)  to obtain global topic hierarchies to improve retrieval performance. In this paper, 
we encode the corpus and the user's search history for a global state representation.

Another related area to our work is dimension reduction. 
Many breakthroughs in neural models for Natural Language Processing (NLP) are built upon word2vec \cite{DBLP:conf/nips/MikolovSCCD13} and its derivation doc2vec \cite{DBLP:conf/icml/LeM14}. Doc2vec is able to transform a high-dimensional discrete text representation into low-dimensional continuous vectors.  Unfortunately, however, 
doc2vec cannot solve a  problem known as \textit{crowding} \cite{DBLP:journals/jmlr/CookSMH07}.  It refers to the situation where multiple high-dimensional data points are collapsed into one after dimension reduction and two data points belonging to different classes are then inseparable. In our case, each data point represents either a relevant or irrelevant document. We choose to use the t-Distributed Stochastic Neighbor Embedding (t-SNE) method \cite{tsne}. It was used to support data visualization for high-dimensional images \cite{tsne}, network parameters \cite{mnih2015human} and word vectors \cite{li2015visualizing}. 
By assuming a t-distribution for the post-reduction distribution, t-SNE provides more space to scatter  data points that were supposed to be collapsed and the dimensions  can be reduced from thousands to as low as 2 or 3. 
Our experiments (Section \ref{exp}) shows that t-SNE outperforms doc2vec for us.

\section{Dynamic Search Background}

Dynamic search (DS) systems are multi-turn interactive agents that assist human users for goal-oriented information seeking~\cite{yang2016dynamic}. DS shares similar traits with its sister AI applications such as task-oriented dialogue systems and multi-turn QA systems. These traits include (1) a goal-oriented task and (2) the interactions with a human user. They exhibit different forms of interactions, though. In DS, the form of interaction is  querying and retrieving  documents. In dialogue systems, it is  generating natural language responses. In multi-turn QA, it is questioning and finding answers.

 DS is backed up by a long line of prior research in information science, library science, and information retrieval (IR). It is originated from Information Seeking (IS) \cite{belkin1993interaction}. 
Most IS research has focused on studying user behaviors \cite{daronnat2019human}.
\citeauthor{Luo:2014:WSD:2600428.2609629}  simplified IS into DS by separating the modeling on the system side from that on the user side and emphasized on the first.  In DS, a human user either becomes part of the environment \cite{DBLP:conf/ictir/LuoDY15} or  another agent who also interacts with the environment.   

RL offers natural solutions to DS. The RL agent, a.k.a. the search engine, observes state at time $t$ ($\bm{s}_t$) from  the environment (the user and the document collection) and takes actions ($a_t$) to retrieve documents to show to the user. An immediate reward $r_t$  is encapsulated in the user's feedback, which expresses how much the retrieved documents would satisfy the user's  informational need. The retrieved documents could also change the states and transition from the old state to a new one: $\bm{s}_t \rightarrow \bm{s}_{t+1}$. This process may continue for many iterations until the search task is accomplished or the user decides to abandon it. Model-based RL approaches, such as Markov Decision Processes (MDPs) and Partially Observable Markov Decision Processes (POMDPs), and policy-based RL approaches have been explored in the past \cite{yang2016dynamic}.

The Dynamic Domain (DD) Tracks held at the Text REtrieval Conference (TREC) from 2015 to 2017 \cite{trecdd17} are campaigns that evaluate DS systems. In the DD Tracks,  human users are replaced with a simulator to provide feedback to the DS agents. The simulator's feedback includes ratings to documents returned by the agent and  which passages in those documents are relevant. 
It was created based on ground truth assessment done by third-party human annotators. 

Dozens of teams participated in the DD Tracks. 
Their methods ranged from results diversification \cite{DBLP:conf/trec/MoraesSZZ16}, relevance feedback 
\cite{DBLP:conf/trec/BuccioM16}, imitation learning \cite{xue2014ictnet}, to the focus of this paper, reinforcement learning \cite{tang2017reinforcement,DBLP:conf/emnlp/AissaSD18}.

\section{The Approach} \label{approach}

Most RL-based DS approaches  can be summarized by a general formulation. It takes two steps. First, a new, temporary query is generated by the RL agent's policy $\pi$: 
\begin{equation}
    q' \leftarrow h(a_t, \sigma )
\end{equation}
where $a_t = \pi(\bm{s}_t, \bm{\theta}_t)$ is the action generated by  policy $\pi$ for state  $\bm{s}_t$ at time $t$ and $\pi$ is parameterized by $\bm{\theta}_t$. We call $h$  the {\it query reformulation function},  which constructs the new query $q'$ based on $\pi$ and heuristics $\sigma$. Note that $\sigma$ does not depend on $t$. $h$ is a non-differentiable function.   

Second,  the newly formulated query $q'$ is sent to 
\begin{equation}
    f(q', d_i, \phi)
\end{equation}
to obtain documents that are, hopefully, relevant to $q'$. We call $f$ the {\it retrieval function}. It is usually an existing retrieval method, such as BM25 or Language Modelling, which is non-differentiable. $f$ returns a score to quantify the relevance between $q'$ and (a document) $d_i$. $f$ is parameterized by $\phi$, which is a variable of the retrieval method and independent of $t$.

Overall, at each search iteration $t$, the RL agent assigns a ranking score to the $i^{th}$ document $d_i$:  
\begin{equation} \label{eq:tradition_form2} 
\resizebox{0.6\columnwidth}{!}{$
{Score}_{i,t} = f(h(\pi(\bm{s}_t, \bm{\theta}_t), \sigma), d_i, \phi) 
$}
\end{equation}
and then all documents are ranked and retrieved based on this scoring.

The first issue of this formulation is that the retrieval function $f$ only finds the top relevant documents for query $q'$.   
At any time, the agent is not aware of the global picture of the state space.  This is different from how an RL agent is treated in AI, where the agent is always aware of the global status of the environment, e.g. AlphaGo knows the game board. This inadequate knowledge of the global picture  hampers the RL agent to make longer-term plans for better decision-making. 

The second issue in this formulation is that neither  $h$ nor  $f$ is differentiable. This prevents a  gradient-based RL  method from correctly updating its gradient. The RL agent is thus unable to effectively adjust the retrieval results based on user feedback. In addition, folding together multiple non-differentiable  functions  makes it very difficult to diagnose an failed action, which could result from a bad policy $\pi$, a sloppy $\sigma$,  or an ineffective $\phi$.

In this paper, we propose  to convert and compress a collection of documents into a global representation. We also introduce a novel differentiable ranking function that can be easily incorporated into a gradient-based RL method \cite{ppo} to perform dynamic search. 


\subsection{Algorithm Overview}

\begin{figure}[t]
\includegraphics[width=0.9\columnwidth]{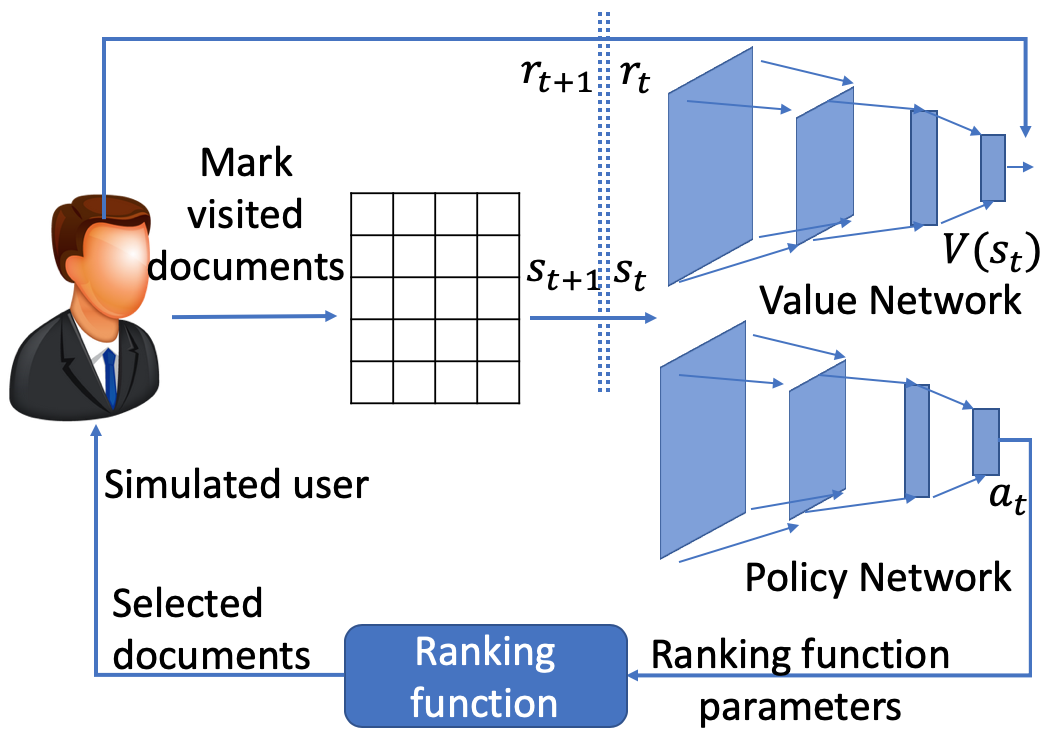} 
\caption{System architecture.}\label{fig:system}  
\end{figure}


Our framework is based on a state-of-the-art Monte Carlo policy-gradient method, Proximal Policy Optimization (PPO) \cite{ppo}. The RL agent consists of two networks, a value network and a policy network. Given the current state $\bm{s}_t$,  the policy network outputs action $\bm{a}_t^{\bm{\theta}}$ and the value network outputs state-value $V(\bm{s}_t)$. 
Both networks  are composed of a few layers of Convolutional Neural Networks (CNNs) 
and a Multi-Layer Perceptron (MLP).  They share the same network structure but use different parameter settings. Figure \ref{fig:system} illustrates  the system architecture.

\begin{algorithm}[t]
\SetAlgoLined
 Initialize $\bm{\theta}$\;
 \For{iteration = 1, 2, ...}{
 	\For{t=1,2, ... , T}{
 		read in the global representation $\bm{s}_t$\;
 		sample action $\bm{a}_t^{\bm{\theta}}=\pi (\bm{s}_t, \bm{\theta}_t)$\;
 		\For{i=1,2, ... , C}{
 		    estimate a relevance score for document $d_i$:
 			${score}_{i,t}= f(\bm{a}_t^{\bm{\theta}}, d_i)$ (Eq. \ref{eq:rank_fn})
 		}
 		rank the documents by ${score}_{*,t}$ and return the top-ranked documents $\mathcal{D}_t$\;
        compute $r_t$ based on Eq. \ref{eq:reward}\;
 		mark returned documents as visited, generate next state $\bm{s}_{t+1}$\;
 		Compute advantage $\hat{A}_t$\;
 	}
 	Optimize $L(\bm{\theta})$\ (Eq. \ref{eq:ppo_loss}) w.r.t. $\bm{\theta}$;
 }
 \caption{CE3.} 
 \label{algo}
\end{algorithm}


Algorithm \ref{algo} describes  the  proposed CE3 method. 
It starts by sampling actions 
by the RL agent. The documents are then ranked by their estimated relevance scores calculated based on the action vectors. The top-ranked ones are shown to the user. The user then examines the documents and submits feedback, which is used to derive the immediate reward $r_t$. 
As a Monte Carlo algorithm,  CE3 generates search trajectories by sampling actions based on the current policy. 
Given enough trajectory samples, the following  objective function is optimized w.r.t. the network parameter $\bm{\theta}$: 
\begin{equation}\label{eq:ppo_loss} 
\resizebox{\columnwidth}{!}{$
L(\bm{\theta}) = \hat{\mathbb{E}}_t [L^{Policy}_t(\bm{\theta}) - c_1 L^{Value}_t(\bm{\theta})  + c_2 S[\pi_{\bm{\theta}}](\bm{s}_t) ] 
$}
\end{equation} where $c_1$ is the weight for the value network and $c_2$ is the coefficient for the policy's entropy, which encourages the agent to explore all areas in the action space. 

Learning of the value network is done by minimizing $L_t^{Value}$. It is the mean squared error 
between the estimated state value $V_{\bm{\theta}}(\bm{s}_t)$  and the targeted state value  $V_t^{targ}$ in the sampled trajectories at time $t$:
\begin{equation} 
\resizebox{0.6\columnwidth}{!}{$
L^{Value}_t (\bm{\theta)} = \hat{\mathbb{E}}_t [(V_{\bm{\theta}}(\bm{s}_t) - V_t^{targ})^2]
$}
\end{equation} 


Learning of the policy network is done by minimizing $L^{Policy}$. It is a pessimistic bound of the effectiveness  
of the policy network at time $t$:
\begin{equation} \label{eq:ppo_loss_policy}
\resizebox{\columnwidth}{!}{$
L_t^{Policy}(\bm{\theta}) = \hat{\mathbb{E}}_t [ \min(\rho_t(\bm{\theta}) \hat{A}_t, clip(\rho_t(\bm{\theta}), 1 - \epsilon, 1+ \epsilon) \hat{A}_t)  ]
$}
\end{equation} 
where 
 $\hat{A}_t$ is the advantage function \cite{Sutton1998}.

Certain actions may increase the return in  extreme situations but may not work in general. To avoid such situation,   
 the algorithm  adopts  a surrogate clip function and discards  actions  
 when their rates-to-change are larger than $\epsilon$:   
 \begin{equation}
 \resizebox{\columnwidth}{!}{$
clip(\rho_t(\bm{\theta}), 1 - \epsilon, 1+ \epsilon) \hat{A}_t = \left \{
	\begin{array}{ll}
    \min(\rho_t(\bm{\theta}), 1+\epsilon) A_t & A_t > 0 \\
    \max(\rho_t(\bm{\theta}), 1-\epsilon) A_t & A_t < 0\\
	\end{array}
\right .
$}
\end{equation}
where $
     \rho_t(\bm{\theta}) = \frac{\pi_{\bm{\theta}} (\bm{a}_t | \bm{s}_t)}{\pi_{\bm{\theta}_{old}} (\bm{a}_t | \bm{s}_t)}
 $ is the change rate of actions.

 The algorithm employs stochastic gradient ascent (SGA) to optimize both the  policy network and the value network. The process  continues until no better policy is found. 

This PPO-based method alone can be used in other applications. However, for the reasons  motivating this paper, we think it should be used in combination with what we will present next --  corpus-level document representation and differentiable ranking function. 

\subsection{Build a Global Representation} \label{board}




In this paper, we propose to compress an entire text corpus into a global low-dimensional representation and keep it at all time. Our goal is to  enable a DS agent to always gain access to the full state space. We believe it is essential for a DS agent because not being able to reach documents in under-explored areas  would mean not being able to recover from early bad decisions. 

We summarize the procedure of creating global representation  into three steps. First, each document is split into topic-coherent segments. 
The latest advances in Neural Information Retrieval (NeuIR) have demonstrated the effectiveness of using topical structures for NeuIR  \cite{tang2018deeptilebars,DBLP:conf/sigir/FanGLXZC18}. In this work, we follow \cite{tang2018deeptilebars} for segmenting and standardizing documents. Each document is split into a fixed $B$ number of segments ($B$ is empirically set to 20). Within each segment, the content is expected to be topic-coherent since the segmentation is done based on Tilebars.  Tilebars  \cite{tilebars} is a classical text visualisation work and has been proven to be very effective in helping identify relevant documents by visualizing the term matches.  

Second, bag-of-Words (BoW) is used as the feature vector for a segment and is of a size equal to the vocabulary's size $W$. This dimension is usually quite high and could easily reach millions in natural language tasks. 
Therefore, we compress each segment into a much manageable lower-dimension $n$ ($n \ll W$).  One challenge is that after the compression the relevant and irrelevant documents would be crowed together and difficult to be separated apart. To address this issue, 
We employ t-SNE \cite{tsne} for dimension reduction.   
The idea is based on Barnes-Hut approximation \cite{barnes1986hierarchical}. Assume the high-dimensional input $\bm{x}_* \in \mathds{R}^W$ follows  Gaussian distribution. The probability that two random data points $\bm{x}_i$ and $\bm{x}_j$ are neighboring to each other is  
\begin{equation}
p(\bm{x}_i, \bm{x}_j) = \frac{exp(-||\bm{x}_i - \bm{x}_j||^2 / 2 \sigma^2)}{\sum_{k \neq l} exp(||\bm{x}_k - \bm{x}_l||^2 / 2 \sigma ^2) } 
\end{equation}
The algorithm then maps these data points in the high dimensional space to points $\bm{y}_*$ in  a  much lower dimensional space $\mathds{R}^n$. 
Suppose $\bm{x}_i$ and $\bm{x}_j$  project into the lower dimension as $\bm{y}_i$ and $\bm{y}_j$. The probability that $\bm{y}_i$ and $\bm{y}_j$ are still neighboring to each other is    
 \begin{equation}
 q(\bm{y}_i, \bm{y}_j) = \frac{(1+||\bm{y}_i - \bm{y}_j||^2)^{-1}}{ \sum_{k \neq l} (1+||\bm{y}_k - \bm{y}_l||^2)^{-1}} 
 \end{equation}
To establish the mapping between $\bm{x}$ and $\bm{y}$, the points' KL divergence 
\begin{equation}
L_{tsne} (\bm{y}||\bm{x}) = \sum_i \sum_j p(\bm{x}_i, \bm{x}_j) \log \frac{p(\bm{x}_i, \bm{x}_j)}{q(\bm{y}_i, \bm{y}_j)}
\end{equation}
is minimized. The solution to the new projection can be achieved step by step via gradient descent. 

\begin{figure}[t]
\centering
\includegraphics[width=0.8\columnwidth]{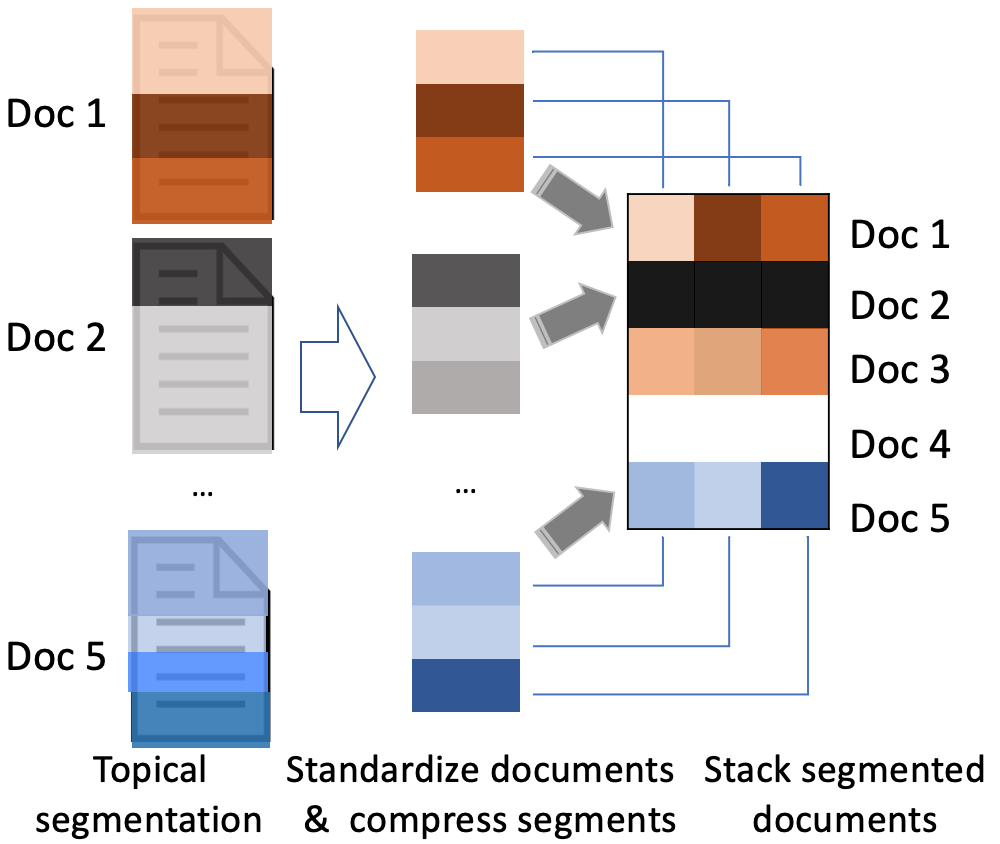} 
\caption{Global representation of a toy corpus (of 5 documents): Documents are segmented and standardized following \cite{tang2018deeptilebars}. Similar colors suggest similar contents. Document 2 is darkened after being visited.  Document 4 is currently selected by the RL agent and highlighted with white.}\label{fig:board_ex} 
\end{figure}

Third, segments from all documents are stacked together to form a global representation. The global representation is denoted by $\mathcal{C}$ and its dimensions are $C \times B \times n$. Here $C$ is the number of documents, $B$ is the number of segments per document, and $n$ is the reduced feature dimension.  In our work, $n$  is  empirically set to 3.  In this global representation $\mathcal{C}$, each row represents a document and each column represents a segment at a certain position in the documents.  Each row unfolds the segments  horizontally,  with their original order in a document preserved. For generality, we make no assumption about the stacking order of documents. 
The RL agent is expected to complete the search task even when  dealing with randomly ordered documents. 
Figure \ref{fig:board_ex} illustrates the  global representation of a toy corpus.

This global representation constructs the states. Our state at time $t$, $\bm{s}_t$,  has two parts,   $\mathcal{C}$ and the retrieval history of documents from time 1 to  $t-1$: 
\begin {equation} 
\label{eq:state_defn}
   \bm{s}_t = S(\mathcal{C}, \mathcal{D}_1 \cup \mathcal{D}_2 \cup ...  \mathcal{D}_i ... \cup \mathcal{D}_{t-1}) 
\end{equation}
where  $\mathcal{D}_i$ is the set of   documents retrieved at time $i$. 

In Algorithm \ref{algo}, already-retrieved documents are marked as visited. In the global representation, it is done by assigning a reserved value to those documents' feature vectors. When the $i^{th}$  document is visited, the feature vectors of all its segments, i.e. $v_{i*}$, are changed to the reserved value. Such change explicitly shows past search history and exploration status at the corpus level.

\subsection{Retrieve using a Differentiable Ranking Function} 

It is crucial for an RL agent to employ a differentiable ranking function as its action so that it can perform end-to-end retrieval. Unfortunately, most existing DS approaches still use ranking functions that are non-differentiable. 

The existing approaches' formulation is shown in Eq. \ref{eq:tradition_form2} ${Score}_{i,t} = f(h(\pi(\bm{s}_t, \bm{\theta}_t), \sigma), d_i, \phi)$. It is is clearly a non-differentiable function. This prevents the RL agent from directly manipulating documents based on user feedback.  We propose to omit query reformulation $h$ completely, including its heuristic $\sigma$. Since our RL agent  would not use any  conventional retrieval model, their heuristic parameter $\phi$ is gone, too. 
The ranking function then becomes:
\begin{equation}\label{eq:action_defn}
{score}_{i,t} = f( \pi(\bm{s}_t, \bm{\theta}_t), d_i) 
\end{equation}



We then focus on making $f$ differentiable. It is  achieved  by using a linear formulation for $f$.  
In our formulation,  $f$ approximates a document $d_i$ 's  relevance as a linear function over the segments belonging to $d_i$: 
\begin{equation}\label{eq:rank_fn}
	{score}_{i,t} = f(\bm{a}_t^{\bm{\theta}},d_i) =\sum_{j=1}^{B} \bm{y}_{ij} \cdot \bm{a}_t^{\bm{\theta}}
\end{equation}
where $\bm{a}_t^{\bm{\theta}} = \pi(\bm{s}_t, \bm{\theta}_t)$ is the action that is generated by  policy  $\pi(\bm{s}_t, \bm{\theta}_t)$ and $\bm{y}_{ij}$  is the feature vector of the $j^{th}$ segment in  $d_i$ after compression. The action vector $\bm{a}_t^{\bm{\theta}}$ can be sampled 
by the RL agent at each time step.  When it gets updated, the new action $\bm{a}_{t+1}^{\bm{\theta}}$ allows the agent to retrieve and explore a different set of documents.

\subsection {Get Reward based on User Feedback}

In TREC DD \cite{trecdd17}, the relevance ratings are provided by the simulated user. We define the immediate reward  as the accumulated relevance ratings (without  discounting) for the retrieved documents. Duplicated results are excluded.  The immediate reward is: 
\begin{equation}\label{eq:reward}
r_t = \sum_{d_i \in \mathcal{D}_t \backslash (\mathcal{D}_1 \cup \mathcal{D}_2 \cup ... \cup \mathcal{D}_{t-1}) }  rel(d_i)
\end{equation}
where  $rel(*)$  is a rating given by the simulator. The rating  can be a positive number $rel \in (0, +\infty)$ for a relevant document ($d_i \in \mathcal{R^+}$), or $0$ for an irrelevant document ($d_i \in \mathcal{R^-}$). The larger the rating, the better the retrieved document.

\section{Experiments} \label{exp}

\subsection{Experimental Settings}

The Text REtrieval Conference (TREC) Dynamic Domain (DD) Tracks 2015 - 2017 \cite{trecdd17} provides a standard testbed for DS.  A simulated user\footnote{https://github.com/trec-dd/trec-dd-jig} issues a starting query, and then provides feedback for all the subsequent runs of retrievals. The feedback includes graded document-level and passage-level relevance judgments in the scale of -1 to 4. 

We experiment on the TREC DD 2017 Track for its judgements' completeness. TREC DD 2017 used LDC New York Times collections \cite{nyt_corpus} as its corpus. The collection included more than 1.8 million news articles archived in the past 20 years. The Track released 60 search tasks created by human assessors. Each task consisted of multiple hierarchically-organized subtopics. The subtopics were not made available to the participating DS systems.  Instead of  post-submission pooling, the Track spent a great deal of efforts in obtaining a complete set of relevant passages before the evaluation started. These answers were used to generate feedback by the simulator. In total, 194 subtopics and 3,816 relevant documents were curated.

Table \ref{tab:topic_ex} shows an example DD search topic DD17-10. In this example, the search task is to find relevant information on `` closing of Leaning Tower in Pisa". 
Table \ref{tab:interaction_ex} shows an example interaction history.

\subsubsection{Metrics}

The evaluation in DS focuses on gaining relevant information  throughout the {\it whole} process. We adopt multiple metrics to evaluate the approaches from various perspectives. \textbf{Aspect recall} \cite{lagergren1998comparing} measures subtopic  coverage: $AsepctRecall = \frac{\mbox{\# subtopics found}}{\mbox{\# subtopics in the topic}}$. 
\textbf{Precision}  and \textbf{Recall} measure the ratios of correctly retrieved documents over the retrieved document set or the entire correct set, respectively: 
$Precision = \frac{|( \cup_{i=1}^n \mathcal{D}_i ) \cap \mathcal{R}^+ |}{ |\cup_{i=1}^n \mathcal{D}_i |  }$, and $Recall= \frac{ |(\cup_{i=1}^n \mathcal{D}_i)  \cap \mathcal{R}^+| }{ |\mathcal{R}^+|} $. 
\textbf{Normalized Session Discounted Cumulative Gain (nsDCG)} evaluates the graded relevance for a ranked document list, putting heavier weights on the early retrieved ones \cite{DBLP:conf/ecir/JarvelinPDN08}:  $sDCG = \sum_{i=1}^n \sum_{d_j \in \mathcal{D}_i} rel(d_j) \left( (1+\log_b j) (1+\log_{bq} i)  \right)^{-1} $, and $nsDCG = \frac{sDCG}  {ideal\ sDCG}$. 

\begin{table}[t]
    \centering
    \resizebox{1\columnwidth}{!}{
    \begin{tabular}{cc}
        \toprule
        Topic (DD17-10)  &   Leaning Towers of Pisa Repairs  \\
        \midrule
        Subtopic 1 (id: 321)& Tourism impact of repairs/closing \\
        Subtopic 2 (id: 319)& Repairs and plans \\
        Subtopic 3 (id: 320)& Goals for future of the tower \\
        Subtopic 4 (id: 318)& Closing of tower\\
        \bottomrule
    \end{tabular}
    }
    \caption{Example Search Topic.}
    \label{tab:topic_ex}
\end{table}

\begin{table}[t]
    \centering
    \begin{tabular}[t]{ll}
        \toprule
    & Search DD17-10 \\ 
    \toprule
        User: & \makecell[tl]{\textit{    Leaning Towers of Pisa Repairs}}\\
        System: & Return document 0290537\\ 
        User: & Non-relevant document. \\
        System: & Return document 0298897\\
        User: & \makecell[tl]{Relevant on subtopic 320 with a rating of 2,\\ \textit{``No one doubts that it will collapse one} \\
        \textit{day unless preventive measures are taken.''}} \\
        System: & Return document 0984009 \\
        User: & \makecell[tl]{Relevant on subtopic 318 with a rating of 4, \\
        \textit{``The 12th-century tower was closed to }\\ 
        \textit{tourists in 1990 for fear it might topple.''}}\\
    \bottomrule
    \end{tabular}
    \caption{Example Interaction History.} 
    \label{tab:interaction_ex}
\end{table}

\subsubsection{Systems}

We compare CE3 to the most recent DS systems.  They were from the TREC DD 2017 submissions. We pick the top submitted run from each team to best represent their approach. 
The runs are:

\textbf{Galago} \cite{DBLP:books/daglib/0022709}:
This approach does not use any user feedback. Documents are repeatedly retrieved with the same query $Q$ at each iteration by Galago. Documents appeared in  previous iterations are removed from  current iteration. 

\textbf{Deep Q-Network (DQN)} \cite{tang2017reinforcement}: A DQN-based algorithm that  selects query reformulation actions such as adding terms and removing terms and uses Galago to retrieve the documents. 

\textbf{Relevance Feedback (RF)} \cite{DBLP:conf/trec/RogersO17} : 
The query $Q$ is used to first retrieve an initial set of documents using Indri.\footnote{https://www.lemurproject.org/indri/} 
Then the documents are re-ranked by their similarity to the user feedback in all previous iterations. It is a variant of the relevance feedback (RF) model \cite{DBLP:journals/jasis/RobertsonJ76}.  

\textbf{Results Diversification (DIV)} \cite{DBLP:conf/trec/ZhangHJWZFYXYLC17}: 
This approach expands queries based on previous user feedback. The  documents retrieved with solr\footnote{http://lucene.apache.org/solr/} are then re-ranked with the xQuAD result diversification algorithm \cite{santos2010explicit}. 

\textbf{CE3}: The proposed method in this paper. For comparison, we also implement a variant, \textbf{CE3 (doc2vec)}, which uses doc2vec \cite{DBLP:conf/icml/LeM14} to compress the feature vector for each segment. The embeddings are trained on more than 1.8 million documents. Other settings are identical between CE3 and CE3 (doc2vec).

\subsubsection{Parameters}

We construct a collection for each search topic $Q$ by mixing relevant documents and irrelevant documents at a ratio of 1:1 to simulate a common re-ranking scenario.  The corpus size $C$ ranges from tens to thousands.  Among all the parameter combinations, the following configuration yields the best performance: The dimension of t-SNE's output $n$ is set to 3. The number of segments per document $B$ is set to 20. Coefficients $c_1$ and $c_2$ in Eq.~\ref{eq:ppo_loss} are $0.5$ and $0$, respectively. 
Both the policy and value networks have 2 layers of CNNs and 1 MLP. The first CNN consists of eight $2 \times 2$ kernels and the second  consists of 16. The hidden layer of MLP consists of 32 units and is the same for both networks. The output layer of MLP has 3 units for the policy network  and 1 for the  value network.

\subsection{Results}


\begin{figure*}[t]
\centering
\includegraphics[width=\textwidth]{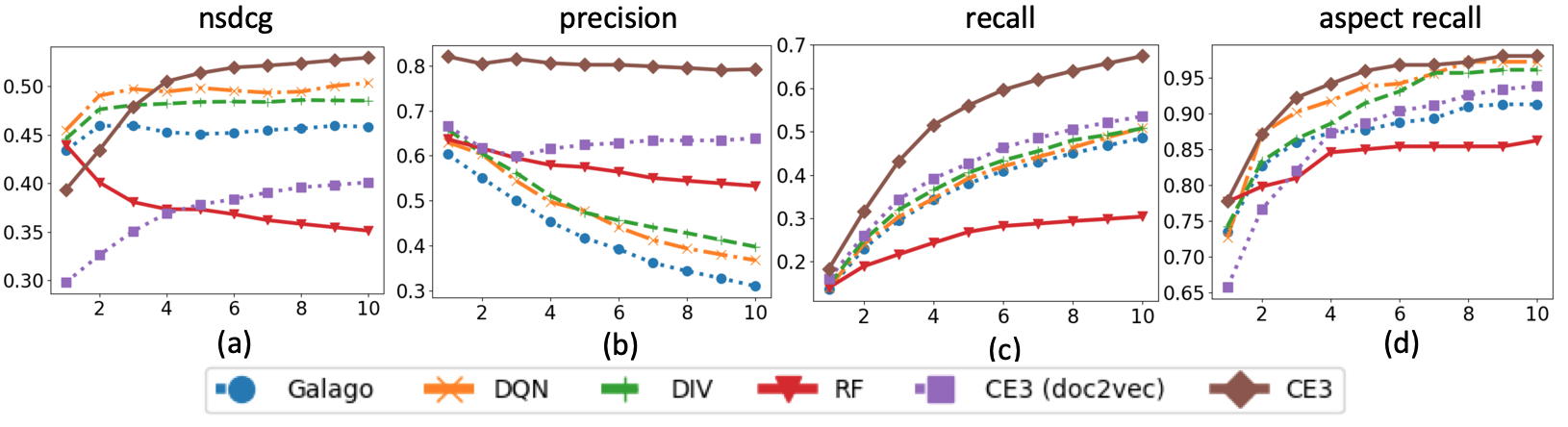} 
\caption{Experiment results in the first 10 search iterations.}\label{fig:results} 
\end{figure*}

\begin{table*}[t]
    \centering
    \resizebox{1.0\textwidth}{!}{
    \begin{tabular}{c|cccccccccc}
    \toprule
         Time step & t=1 &	t=2	&t=3&t=	4	&t=5&	t=6&	t=7&t=8&	t=9&	t=10 \\
    \midrule
         CE3 (doc2vec)& 0.0\% & 11.7\% & 18.0\%	 & 30.0\% &	35.0\% &	34.3\% &	30.0\% &	25.0\% &	25.7\% &	25.0\%  \\
    \hline
         CE3 & 0.0\% & 3.0\% & 1.7\% & 1.0\% & 3.0\% &	0.0\% &	1.0\% &	0.3\% &	3.0\% &	0.0\% \\
    \bottomrule
    \end{tabular}
    }
    \caption{Percentage of duplicate documents.}
    \label{tab:duplicate}
\end{table*}

From Figure \ref{fig:results}, we observe that
 CE3 outperforms all others in recall (Fig. \ref{fig:results}c) and aspect recall (Fig. \ref{fig:results}c) at all time. 
 It suggests that our RL agent is able to explore more areas in the state and action spaces than the rest. 
  While other algorithms also manage to achieve a high aspect recall ($>0.9$), they do not perform as well at recall. It shows that although  traditional diversification methods can find a few relevant documents for each  aspect, it is hard for them to continue the investigation on a visited aspect. This indicates their less effective exploration. 
Instead, CE3's ranking function enables end-to-end optimization, which allows the agent to effectively explore  at all different directions in the state and action spaces. It thus works very well on recall-oriented measures.

CE3 performs very impressive in precision (Fig.  \ref{fig:results}b), too. As the search episode develops, all other approaches show declined precision; however, CE3 stays strong at all iterations. We think it is because other methods could not easily recover from early mistakes while CE3's global representation allows it to explore elsewhere  for ne opportunities when a bad decision happens.

Moreover, even not specifically designed for rank-sensitive metrics, CE3 performs very well on nsDCG, too. Results (Fig.  \ref{fig:results}a)   
reveal that at the beginning CE3 does not score as high as other methods; however, at the end of the episode,  CE3 largely outperforms the rest. We believe the initial successes of  other methods are  caused by that  they are well-tuned to be ranking-sensitive, which is what existing retrieval functions address. However, they seem not to be able to adapt well when the number of interactions increases.

In addition, it comes to our attention that CE3 (doc2vec) is left far behind by CE3. We know that they only differ in their choices to dimension reduction. In a follow-up investigation, we discover that CE3 retrieves much less duplicated documents than CE3 (doc2vec) does. Table \ref{tab:duplicate} reports  $\frac{ | \mathcal{D}_t \cap (\mathcal{D}_1 \cup \mathcal{D}_2 \cup ... \cup \mathcal{D}_{t-1}) |}{|\mathcal{D}_t |}$, the percentage of duplicate documents being retrieved, for the two CE3 variants. We believe it is due to how they compress the feature vectors in a segment. Doc2vec makes no assumption about the data distribution after compression. Vectors trained by doc2vec are probably crowded together and yield more duplicated  results. 
On the contrary, t-SNE helps CE3  separate relevant documents from  irrelevant documents, which makes it contribution to CE's success.

\subsection{Visualize the Exploration}\label{case_study}
\begin{figure}[t]
\centering
\includegraphics[width=\columnwidth]{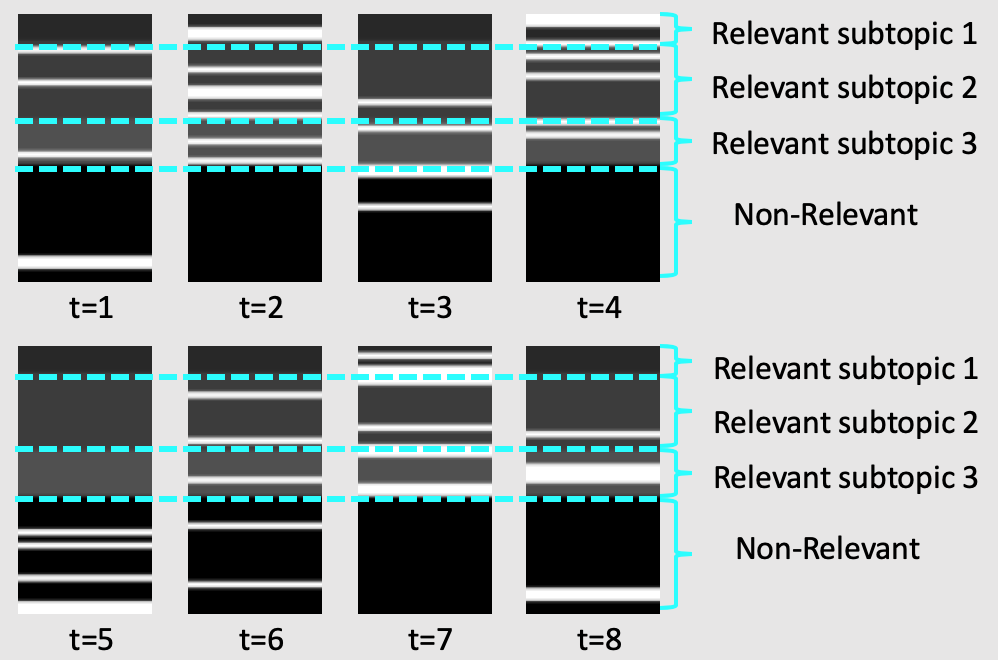} 
\caption{
Visualization of Exploration (Topic DD17-3). White bars mark documents selected by the agent. Successful retrieval means more white in the top half.}\label{fig:exploration}
\end{figure}

 We are interested in observing the dynamics during a DS process.  Figure \ref{fig:exploration} illustrates the first 8  steps for a search task with 3 subtopics. 
Based on the ground truth, we arrange the relevant documents  at the top and  irrelevant documents  at the bottom. Among the relevant documents, those belong to the same subtopic are grouped together and placed in the order of subtopics 1 to 3. The turquoise dotted lines are added to highlight where each subtopic's are. The white color does not indicate relevance but  show the visitations. It highlights  which documents the agent returns at time $t$. A thicker white bar indicates more selected documents in the same subtopic.
In the case a selected document is relevant to multiple subtopics, it is highlighted in multiple places. The effectiveness of the DS agent is jointly told by the white highlights and their positions: {\it successful retrieval means more white in the top half of this picture}.

We observe that at $t=4$,  the DS agent explores at subtopics 1 and 2; while at $t=7$, it changes to subtopics 1 and 3. At the $5^{th}$ iteration, 
 the agent seems to enter into a wrong path since the visualization shows that its current selection is in the lower irrelevant portion. However, the agent quickly corrects its actions and improves at $t=7$ and $t=8$. It confirms  that CE3 is able to recover from bad actions.

\section{Conclusion } \label{conclusion}
Using Dynamic Search (DS) as an illustrating example, this paper presents a new deep reinforcement learning framework for retrieval-based interactive AI systems. To allow an agent to explore a space fully and freely, we propose to maintain a global representation of the entire corpus at all time. We achieve corpus-level compression by t-SNE dimension reduction.  We also propose a novel differentiable ranking function to ensure user feedback can truly control what documents to return. The experimental results demonstrate that our method's  performance is superior to state-of-the-art DS systems. Given the fundamental issues we  address in this paper,  we believe CE3's success can be extended to other interactive AI systems.

\section*{Acknowledgements}
This research was supported by U.S. National Science Foundation IIS-1453721. Any opinions, findings, conclusions, or recommendations expressed in this paper are of the authors, and do not necessarily reflect those of the sponsor.

\bibliography{reference_short}

\begin{thebibliography}{}

\bibitem[\protect\citeauthoryear{Aissa, Soulier, and
  Denoyer}{2018}]{DBLP:conf/emnlp/AissaSD18}
Aissa, W.; Soulier, L.; and Denoyer, L.
\newblock 2018.
\newblock A reinforcement learning-driven translation model for search-oriented
  conversational systems.
\newblock In {\em The 2nd International Workshop on Search-Oriented
  Conversational AI, {SCAI}@{EMNLP} '18}.

\bibitem[\protect\citeauthoryear{Barnes and Hut}{1986}]{barnes1986hierarchical}
Barnes, J., and Hut, P.
\newblock 1986.
\newblock A hierarchical o (n log n) force-calculation algorithm.
\newblock {\em Nature} 324(6096):446.

\bibitem[\protect\citeauthoryear{Belkin and
  others}{1993}]{belkin1993interaction}
Belkin, N.~J., et~al.
\newblock 1993.
\newblock Interaction with texts: Information retrieval as information seeking
  behavior.
\newblock {\em Information Retrieval} 93:55--66.

\bibitem[\protect\citeauthoryear{Buccio and
  Melucci}{2016}]{DBLP:conf/trec/BuccioM16}
Buccio, E.~D., and Melucci, M.
\newblock 2016.
\newblock Evaluation of a feedback algorithm inspired by quantum detection for
  dynamic search tasks.
\newblock In {\em {TREC} '16}.

\bibitem[\protect\citeauthoryear{Cook \bgroup et al\mbox.\egroup
  }{2007}]{DBLP:journals/jmlr/CookSMH07}
Cook, J.; Sutskever, I.; Mnih, A.; and Hinton, G.~E.
\newblock 2007.
\newblock Visualizing similarity data with a mixture of maps.
\newblock In {\em {AISTATS} '07}.

\bibitem[\protect\citeauthoryear{Croft, Metzler, and
  Strohman}{2009}]{DBLP:books/daglib/0022709}
Croft, W.~B.; Metzler, D.; and Strohman, T.
\newblock 2009.
\newblock {\em Search Engines - Information Retrieval in Practice}.
\newblock Pearson Education.

\bibitem[\protect\citeauthoryear{Daronnat \bgroup et al\mbox.\egroup
  }{2019}]{daronnat2019human}
Daronnat, S.; Azzopardi, L.; Halvey, M.; and Dubiel, M.
\newblock 2019.
\newblock Human-agent collaborations: trust in negotiating control.
\newblock In {\em {CHI} '19}.

\bibitem[\protect\citeauthoryear{Dhingra \bgroup et al\mbox.\egroup
  }{2017}]{P17-1045}
Dhingra, B.; Li, L.; Li, X.; Gao, J.; Chen, Y.; Ahmed, F.; and Deng, L.
\newblock 2017.
\newblock Towards end-to-end reinforcement learning of dialogue agents for
  information access.
\newblock In {\em {ACL} '17}.

\bibitem[\protect\citeauthoryear{Fan \bgroup et al\mbox.\egroup
  }{2018}]{DBLP:conf/sigir/FanGLXZC18}
Fan, Y.; Guo, J.; Lan, Y.; Xu, J.; Zhai, C.; and Cheng, X.
\newblock 2018.
\newblock Modeling diverse relevance patterns in ad-hoc retrieval.
\newblock In {\em {SIGIR} '18}.

\bibitem[\protect\citeauthoryear{Hearst}{1995}]{tilebars}
Hearst, M.~A.
\newblock 1995.
\newblock Tilebars: visualization of term distribution information in full text
  information access.
\newblock In {\em CHI '95}.

\bibitem[\protect\citeauthoryear{Hu \bgroup et al\mbox.\egroup
  }{2018}]{Hu:2018:RLR:3219819.3219846}
Hu, Y.; Da, Q.; Zeng, A.; Yu, Y.; and Xu, Y.
\newblock 2018.
\newblock Reinforcement learning to rank in e-commerce search engine:
  Formalization, analysis, and application.
\newblock In {\em {SIGKDD} '18}.

\bibitem[\protect\citeauthoryear{Huang and
  Efthimiadis}{2009}]{huang2009analyzing}
Huang, J., and Efthimiadis, E.~N.
\newblock 2009.
\newblock Analyzing and evaluating query reformulation strategies in web search
  logs.
\newblock In {\em {CIKM} '09}.

\bibitem[\protect\citeauthoryear{J{\"{a}}rvelin \bgroup et al\mbox.\egroup
  }{2008}]{DBLP:conf/ecir/JarvelinPDN08}
J{\"{a}}rvelin, K.; Price, S.~L.; Delcambre, L. M.~L.; and Nielsen, M.~L.
\newblock 2008.
\newblock Discounted cumulated gain based evaluation of multiple-query {IR}
  sessions.
\newblock In {\em {ECIR} '08}.

\bibitem[\protect\citeauthoryear{Lagergren and
  Over}{1998}]{lagergren1998comparing}
Lagergren, E., and Over, P.
\newblock 1998.
\newblock Comparing interactive information retrieval systems across sites: The
  {TREC-6} interactive track matrix experiment.
\newblock In {\em {SIGIR} '98}.

\bibitem[\protect\citeauthoryear{Le and Mikolov}{2014}]{DBLP:conf/icml/LeM14}
Le, Q.~V., and Mikolov, T.
\newblock 2014.
\newblock Distributed representations of sentences and documents.
\newblock In {\em {ICML} '14}.

\bibitem[\protect\citeauthoryear{Li \bgroup et al\mbox.\egroup
  }{2016}]{li2015visualizing}
Li, J.; Chen, X.; Hovy, E.~H.; and Jurafsky, D.
\newblock 2016.
\newblock Visualizing and understanding neural models in {NLP}.
\newblock In {\em {NAACL} '16}.

\bibitem[\protect\citeauthoryear{Li, Resnick, and Mei}{2016}]{li2016multiple}
Li, C.; Resnick, P.; and Mei, Q.
\newblock 2016.
\newblock Multiple queries as bandit arms.
\newblock In {\em {CIKM} '16}.

\bibitem[\protect\citeauthoryear{Liu and Croft}{2004}]{liu2004cluster}
Liu, X., and Croft, W.~B.
\newblock 2004.
\newblock Cluster-based retrieval using language models.
\newblock In {\em {SIGIR} '04}.

\bibitem[\protect\citeauthoryear{Luo, Dong, and
  Yang}{2015}]{DBLP:conf/ictir/LuoDY15}
Luo, J.; Dong, X.; and Yang, H.
\newblock 2015.
\newblock Session search by direct policy learning.
\newblock In {\em {ICTIR} '15}.

\bibitem[\protect\citeauthoryear{Luo, Zhang, and
  Yang}{2014}]{Luo:2014:WSD:2600428.2609629}
Luo, J.; Zhang, S.; and Yang, H.
\newblock 2014.
\newblock Win-win search: dual-agent stochastic game in session search.
\newblock In {\em {SIGIR} '14}.

\bibitem[\protect\citeauthoryear{Maaten and Hinton}{2008}]{tsne}
Maaten, L. v.~d., and Hinton, G.
\newblock 2008.
\newblock Visualizing data using t-sne.
\newblock {\em JMLR} 9(Nov):2579--2605.

\bibitem[\protect\citeauthoryear{Marchionini}{2006}]{Marchionini:2006:ESF:1121949.1121979}
Marchionini, G.
\newblock 2006.
\newblock Exploratory search: From finding to understanding.
\newblock {\em ACM Communications} 49(4):41--46.

\bibitem[\protect\citeauthoryear{Mikolov \bgroup et al\mbox.\egroup
  }{2013}]{DBLP:conf/nips/MikolovSCCD13}
Mikolov, T.; Sutskever, I.; Chen, K.; Corrado, G.~S.; and Dean, J.
\newblock 2013.
\newblock Distributed representations of words and phrases and their
  compositionality.
\newblock In {\em {NIPS} '13}.

\bibitem[\protect\citeauthoryear{Mnih \bgroup et al\mbox.\egroup
  }{2015}]{mnih2015human}
Mnih, V.; Kavukcuoglu, K.; Silver, D.; Rusu, A.~A.; Veness, J.; Bellemare,
  M.~G.; Graves, A.; Riedmiller, M.; Fidjeland, A.~K.; Ostrovski, G.; et~al.
\newblock 2015.
\newblock Human-level control through deep reinforcement learning.
\newblock {\em Nature} 518(7540):529.

\bibitem[\protect\citeauthoryear{Moraes, Santos, and
  Ziviani}{2016}]{DBLP:conf/trec/MoraesSZZ16}
Moraes, F.; Santos, R. L.~T.; and Ziviani, N.
\newblock 2016.
\newblock {UFMG} at the {TREC} 2016 dynamic domain track.
\newblock In {\em {TREC} '16}.

\bibitem[\protect\citeauthoryear{Reddy, Chen, and
  Manning}{2019}]{DBLP:journals/tacl/ReddyCM19}
Reddy, S.; Chen, D.; and Manning, C.~D.
\newblock 2019.
\newblock Coqa: {A} conversational question answering challenge.
\newblock {\em {TACL}} 7:249--266.

\bibitem[\protect\citeauthoryear{Robertson and
  Jones}{1976}]{DBLP:journals/jasis/RobertsonJ76}
Robertson, S.~E., and Jones, K.~S.
\newblock 1976.
\newblock Relevance weighting of search terms.
\newblock {\em {JASIS}} 27(3):129--146.

\bibitem[\protect\citeauthoryear{Robertson, Zaragoza, and others}{2009}]{bm25}
Robertson, S.; Zaragoza, H.; et~al.
\newblock 2009.
\newblock The probabilistic relevance framework: Bm25 and beyond.
\newblock {\em Foundations and Trends{\textregistered} in Information
  Retrieval} 3(4):333--389.

\bibitem[\protect\citeauthoryear{Rogers and
  Oard}{2017}]{DBLP:conf/trec/RogersO17}
Rogers, K., and Oard, D.~W.
\newblock 2017.
\newblock Umd{\_}clip: Using relevance feedback to find diverse documents for
  {TREC} dynamic domain 2017.
\newblock In {\em {TREC} '17}.

\bibitem[\protect\citeauthoryear{Salton and
  Buckley}{1988}]{DBLP:journals/ipm/SaltonB88}
Salton, G., and Buckley, C.
\newblock 1988.
\newblock Term-weighting approaches in automatic text retrieval.
\newblock {\em Inf. Process. Manage.} 24(5):513--523.

\bibitem[\protect\citeauthoryear{Sandhaus}{2008}]{nyt_corpus}
Sandhaus, E.
\newblock 2008.
\newblock The new york times annotated corpus.
\newblock {\em Linguistic Data Consortium, Philadelphia} 6(12):e26752.

\bibitem[\protect\citeauthoryear{Sankar \bgroup et al\mbox.\egroup
  }{2019}]{DBLP:conf/acl/SankarSPCB19}
Sankar, C.; Subramanian, S.; Pal, C.; Chandar, S.; and Bengio, Y.
\newblock 2019.
\newblock Do neural dialog systems use the conversation history effectively? an
  empirical study.
\newblock In {\em {ACL} '19}.

\bibitem[\protect\citeauthoryear{Santos \bgroup et al\mbox.\egroup
  }{2010}]{santos2010explicit}
Santos, R. L.~T.; Peng, J.; Macdonald, C.; and Ounis, I.
\newblock 2010.
\newblock Explicit search result diversification through sub-queries.
\newblock In {\em {ECIR} '10}.

\bibitem[\protect\citeauthoryear{Schulman \bgroup et al\mbox.\egroup
  }{2017}]{ppo}
Schulman, J.; Wolski, F.; Dhariwal, P.; Radford, A.; and Klimov, O.
\newblock 2017.
\newblock Proximal policy optimization algorithms.
\newblock {\em arXiv preprint arXiv:1707.06347}.

\bibitem[\protect\citeauthoryear{Sutton and Barto}{2018}]{Sutton1998}
Sutton, R.~S., and Barto, A.~G.
\newblock 2018.
\newblock {\em Reinforcement Learning: An Introduction}.
\newblock The MIT Press, second edition.

\bibitem[\protect\citeauthoryear{Tang and Yang}{2017}]{tang2017reinforcement}
Tang, Z., and Yang, G.~H.
\newblock 2017.
\newblock A reinforcement learning approach for dynamic search.
\newblock In {\em {TREC} '17}.

\bibitem[\protect\citeauthoryear{Tang and Yang}{2019}]{tang2018deeptilebars}
Tang, Z., and Yang, G.~H.
\newblock 2019.
\newblock Deeptilebars: Visualizing term distribution for neural information
  retrieval.
\newblock In {\em {AAAI} '19}.

\bibitem[\protect\citeauthoryear{Wei and Croft}{2006}]{wei2006lda}
Wei, X., and Croft, W.~B.
\newblock 2006.
\newblock Lda-based document models for ad-hoc retrieval.
\newblock In {\em {SIGIR} '06}.

\bibitem[\protect\citeauthoryear{White and Roth}{2009}]{white2009exploratory}
White, R.~W., and Roth, R.~A.
\newblock 2009.
\newblock Exploratory search: Beyond the query-response paradigm.
\newblock {\em Synthesis lectures on information concepts, retrieval, and
  services} 1(1):1--98.

\bibitem[\protect\citeauthoryear{Xue \bgroup et al\mbox.\egroup
  }{2014}]{xue2014ictnet}
Xue, Y.; Cui, G.; Yu, X.; Liu, Y.; and Cheng, X.
\newblock 2014.
\newblock {ICTNET} at session track {TREC2014}.
\newblock In {\em {TREC} '14}.

\bibitem[\protect\citeauthoryear{Yang, Sloan, and Wang}{2016}]{yang2016dynamic}
Yang, G.~H.; Sloan, M.; and Wang, J.
\newblock 2016.
\newblock Dynamic information retrieval modeling.
\newblock {\em Synthesis Lectures on Information Concepts, Retrieval, and
  Services} 8(3):1--144.

\bibitem[\protect\citeauthoryear{Yang, Tang, and Soboroff}{2017}]{trecdd17}
Yang, G.~H.; Tang, Z.; and Soboroff, I.
\newblock 2017.
\newblock {TREC} 2017 dynamic domain track overview.
\newblock In {\em {TREC} '17}.

\bibitem[\protect\citeauthoryear{Zhang \bgroup et al\mbox.\egroup
  }{2017}]{DBLP:conf/trec/ZhangHJWZFYXYLC17}
Zhang, W.; Hu, Y.; Jia, R.; Wang, X.; Zhang, L.; Feng, Y.; Yu, S.; Xue, Y.; Yu,
  X.; Liu, Y.; and Cheng, X.
\newblock 2017.
\newblock {ICTNET} at {TREC} 2017 dynamic domain track.
\newblock In {\em {TREC} '17}.

\end{thebibliography}
\bibliographystyle{aaai}
\end{document}